\documentclass[iop,a4paper]{emulateapj}
\usepackage[english]{babel}
\usepackage{graphicx}
\usepackage{amsthm}
\usepackage{bm}
\usepackage{booktabs}
\usepackage{color}
\usepackage{hyperref}
\usepackage{latexsym}
\usepackage{natbib}
\usepackage{times}
\usepackage{stmaryrd}
\newcommand{\ie}{\textit{i.e.}~}

\begin{document}
\title{Universality and intermittency in relativistic turbulent flows of
a hot plasma}
\author{David Radice}
\affil{Max-Planck-Institut f\"ur Gravitationsphysik, Albert Einstein
Institut, Potsdam, Germany}
\email{david.radice@aei.mpg.de}
\author{Luciano Rezzolla}
\affil{Max-Planck-Institut f\"ur Gravitationsphysik, Albert Einstein
Institut, Potsdam, Germany}
\affil{Department of Physics and Astronomy, Louisiana State
University, Baton Rouge, USA}
\begin{abstract}
  With the aim of determining the statistical properties of
  relativistic turbulence and unveiling novel and non-classical
  features, we present the results of direct numerical simulations of
  driven turbulence in an ultrarelativistic hot plasma using
  high-order numerical schemes. We study the statistical properties of
  flows with average Mach number ranging from $\sim 0.4$ to $\sim 1.7$
  and with average Lorentz factors up to $\sim 1.7$. We find that flow
  quantities, such as the energy density or the local Lorentz factor,
  show large spatial variance even in the subsonic case as
  compressibility is enhanced by relativistic effects. The velocity
  field is highly intermittent, but its power-spectrum is found to be
  in good agreement with the predictions of the classical theory of
  Kolmogorov. Overall, our results indicate that relativistic effects
  are able to significantly enhance the intermittency of the flow and
  affect the high-order statistics of the velocity field, while
  leaving unchanged the low-order statistics, which instead appear to
  be universal and in good agreement with the classical Kolmogorov
  theory. To the best of our knowledge, these are the most accurate
  simulations of driven relativistic turbulence to date.
\end{abstract}
\maketitle
\section{Introduction}
Turbulence is an ubiquitous phenomenon in nature as it plays a
fundamental role in shaping the dynamics of systems ranging from the
mixture of air and oil in a car engine, up to the rarefied hot plasma
composing the intergalactic medium. Relativistic hydrodynamics is a
fundamental ingredient in the modeling of a number of systems
characterized by high Lorentz-factor flows, strong gravity or
relativistic temperatures. Examples include the early Universe,
relativistic jets, gamma-ray-bursts (GRBs), relativistic heavy-ion
collisions and core-collapse supernovae~\citep{Font08}.

Despite the importance of relativistic hydrodynamics and the reasonable
expectation that turbulence is likely to play an important role in many
of the systems mentioned above, extremely little is known about
turbulence in a relativistic regime. For this reason, the study of
relativistic turbulence may be of fundamental importance to develop a
quantitative description of many astrophysical systems.  Furthermore the
comparative study of classical and relativistic turbulence can be useful
also for a better understanding of classical turbulence. For instance,
the study by \citet{Cho2005} of relativistic force-free turbulence, \ie
MHD turbulence in the limit where the plasma inertia and momentum are
neglected, gave important insights in the understanding of
strong-Alfv\'enic turbulence. In particular, it provided a first
important confirmation of the model by \citet{Goldreich1995}, whose
prediction of a $-5/3$ slope for the energy spectrum has been recently
confirmed in classical MHD by \citep{Beresnyak2009, Beresnyak2011a}.  To
this aim, we have performed a series of high-order direct numerical
simulations of driven relativistic turbulence of a hot plasma.
\section{Model and method}
We consider an idealized model of an ultrarelativistic fluid with
four-velocity $u^{\mu} = W (1, v^i)$, where $W \equiv (1 -
v_iv^i)^{-1/2}$ is the Lorentz factor and $v^i$ is the three-velocity
in units where $c = 1$. The fluid is modeled as perfect and described
by the stress-energy tensor 
\begin{equation}
T_{\mu\nu} = (\rho + p) u_{\mu} u_{\nu} + p\, g_{\mu\nu}\,,
\end{equation}
where $\rho$ is the (local-rest-frame) energy density,
$p$ is the pressure, $u_{\mu}$ the four-velocity, and $g_{\mu\nu}$ is
the spacetime metric, which we take to be the Minkowski one. We evolve
the equations describing conservation of energy and momentum in the
presence of an externally imposed Minkowskian force $F^{\mu}$, \ie
$\nabla_{\nu} T^{\mu\nu} = F^{\mu}$, where the forcing term is written
as $F^{\mu} = \tilde{F}(0, f^i)$.  More specifically, the spatial part
of the force, $f^i$, is a zero-average, solenoidal, random, vector
field with a spectral distribution which has compact support in the
low wavenumber part of the Fourier spectrum. Moreover, $f^{i}$, is
kept fixed during the evolution and it is the same for all the models,
while $\tilde{F}$ is either a constant or a simple function of time
(see below for details).

\begin{figure*}
  \begin{minipage}{0.49\hsize}
    \includegraphics[width=\columnwidth]{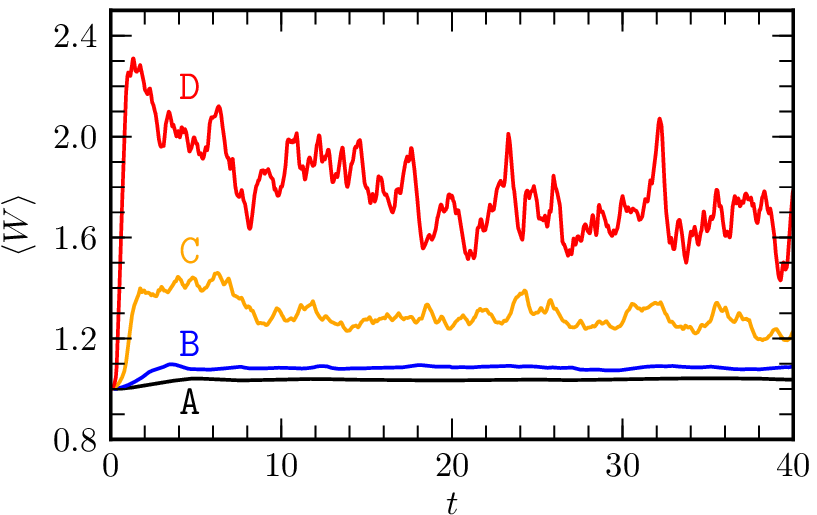}
  \end{minipage}
  \begin{minipage}{0.49\hsize}
  \begin{center}
    \includegraphics[width=0.8\columnwidth]{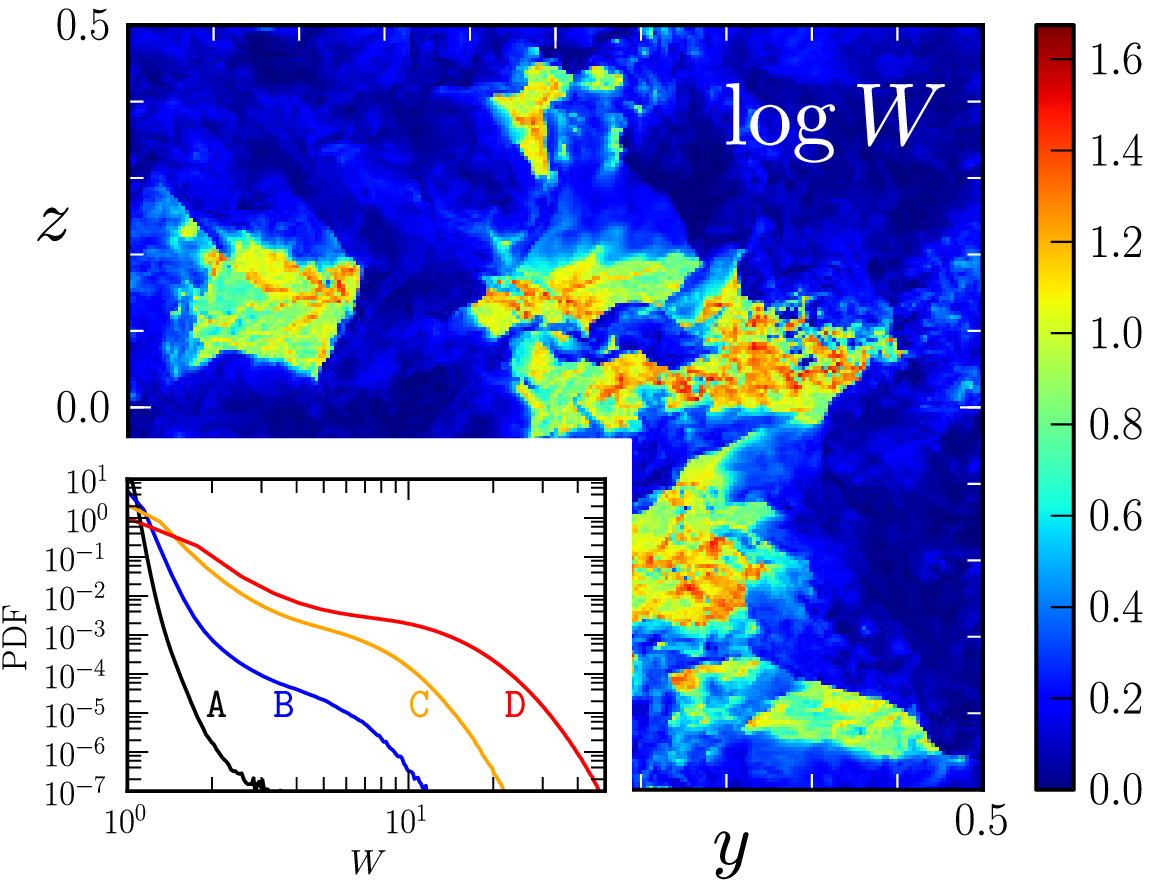}
  \end{center}
  \end{minipage}
  \caption{
  \textit{Left panel:} average Lorentz factor as a function of time for
  the different models considered. Note that a quasi-stationary state is
  reached before $t\sim 10$ for all values of the driving force.
  \textit{Right panel:} logarithm of the Lorentz factor on the $(y,z)$
  plane at the final time of model \texttt{D}. Note the large spatial
  variations of the Lorentz factor with front-like structures. The
  time-averaged PDFs are shown in the lower left corner for the different
  models considered.
  \label{fig:lorentz}
  }
\end{figure*}

The time component of the forcing term, $F^0$, is set to be zero, so that
the driving force is able to accelerate fluid elements without changing
their total energy (in the Eulerian frame). Note that this is
conceptually equivalent to the addition of a cooling term balancing the
effect of the work done on the system by the driving force. On the other
hand, we impose a minimum value for the energy density in the
local-rest-frame, $\rho_{\rm min}$. This choice is motivated essentially
by numerical reasons (the very large Lorentz factor produced can lead to
unphysical point-wise values of $\rho$) and has the effect of slowly
heating up the fluid.  Furthermore, this floor does not affect the
momentum of the fluid and only the temperature is increased.  From a
physical point of view, our approach mimics the fact that in the
low-density regions, the constituents of the plasma are easily
accelerated to very high Lorentz factors, hence emitting bremsstrahlung
radiation heating up the surrounding regions. The net effect is that
energy is subtracted from the driving force and converted into thermal
energy of the fluid, heating it up. In general $\rho_{\rm min}$ is chosen
to be two orders of magnitude smaller than the initial energy density,
but we have verified that the results presented here are insensitive to
the specific value chosen for $\rho_{\rm min}$ by performing simulations
where the floor value is changed by up to two orders of magnitude without
significant differences.

The set of relativistic-hydrodynamic equations is closed by the
equation of state (EOS) $p = \frac{1}{3} \rho$, thus modelling a hot,
optically-thick, radiation-pressure dominated plasma, such as the
electron-positron plasma in a GRB fireball or the matter in the
radiation-dominated era of the early Universe. The EOS used can be
thought as the relativistic equivalent of the classical isothermal EOS
in that the sound speed is a constant, \ie $c_s^2 = 1/3$. At the same
time, an ultrarelativistic fluid is fundamentally different from a
classical isothermal fluid. For instance, its ``inertia'' is entirely
determined by the temperature and the notion of rest-mass density is
lost since the latter is minute (or zero for a pure photon gas) when
compared with the internal one. For these reasons, there is no direct
classical counterpart of an ultrarelativistic fluid and a relativistic
description is needed even for small velocities.

We solve the equations of relativistic hydrodynamics in a 3D periodic
domain using the high-resolution shock capturing scheme described
in~\citep{Radice2012a}. In particular, ours is a flux-vector-splitting
scheme~\citep{Toro99}, using the fifth-order MP5
reconstruction~\citep{suresh_1997_amp}, in local characteristic
variables~\citep{Hawke2001}, with a linearized flux-split algorithm
with entropy and carbuncle fix~\citep{Radice2012a}.
\section{Basic flow properties}
Our analysis is based on the study of four different models, which we
label as \texttt{A}, \texttt{B}, \texttt{C} and \texttt{D}, and which
differ for the initial amplitude of the driving factor $\tilde{F}=1, 2,
5$ for models \texttt{A}--\texttt{C}, and $\tilde{F}(t) = 10 +
\frac{1}{2} t$ for the extreme model \texttt{D}. Each model was evolved
using three different uniform resolutions of $128^3$, $256^3$ and $512^3$
grid-zones over the same unit lengthscale. As a result, model \texttt{A}
is subsonic, model \texttt{B} is transonic and models \texttt{C} and
\texttt{D} are instead supersonic. The spatial and time-averaged
relativistic Mach numbers $\langle v W \rangle/(c_s W_s)$ are $0.362$,
$0.543$, $1.003$ and $1.759$ for our models \texttt{A}, \texttt{B},
\texttt{C} and \texttt{D}, while the average Lorentz factors are $1.038$,
$1.085$, $1.278$ and $1.732$ respectively

The initial conditions are simple: a constant energy density and a
zero-velocity field. The forcing term, which is enabled at time $t = 0$,
quickly accelerates the fluid, which becomes turbulent. By the time when
we start to sample the data, \ie at $t = 10$ (light-)crossing times,
turbulence is fully developed and the flow has reached a stationary
state. The evolution is then carried out up to time $t = 40$, thus
providing data for 15, equally-spaced timeslices over $30$ crossing
times. As a representative indicator of the dynamics of the system, we
show in the left panel of Fig.~\ref{fig:lorentz} the time evolution of
the average Lorentz factor for the different models considered. Note that
the Lorentz factor grows very rapidly during the first few crossing times
and then settles to a quasi-stationary evolution. Furthermore, the
average grows nonlinearly with the increase of the driving term, going
from $\langle W \rangle \simeq 1.04$ for the subsonic model \texttt{A},
up to $\langle W \rangle \simeq 1.73$ for the most supersonic model
\texttt{D}.

Flow quantities such as the energy density, the Mach number or the
Lorentz factor show large spatial variance, even in our subsonic model.
Similar deviations from the average mass density, have been reported also
in classical turbulent flows of weakly compressible
fluids~\citep{Benzi2008}, where it was noticed that compressible effects,
leading to the formation of front-like structures in the density and
entropy fields, cannot be neglected even at low Mach numbers. In the same
way, relativistic effects in the kinematics of the fluid, such those due
to nonlinear couplings via the Lorentz factor~\citep{Rezzolla02}, have to
be taken into account even when the average Lorentz factor is small. The
probability distribution functions (PDFs) of the Lorentz factor are shown
in the right panel of Fig. \ref{fig:lorentz} for the different models.
Clearly, as the forcing is increased, the distribution widens, reaching
Lorentz factors as large as $W\simeq 40$ (\ie to speeds $v\simeq
0.9997$). Even in the most ``classical'' case \texttt{A}, the flow shows
patches of fluid moving at ultrarelativistic speeds. Also shown in Fig.
\ref{fig:lorentz} is the logarithm of the Lorentz factor on the $(y,z)$
plane and at $t=40$ for model \texttt{D}, highlighting the large spatial
variations of $W$ and the formation of front-like structures.
\section{Universality}
As customary in studies of turbulence, we have analyzed the power
spectrum of the velocity field 
\begin{equation}
E_{\boldsymbol{v}}(k) \equiv \frac{1}{2} \int_{|\boldsymbol{k}|=k}
| \hat{\boldsymbol{v}}(\boldsymbol{k}) |^2\, d\boldsymbol{k}\,,
\end{equation}
where $\boldsymbol{k}$ is a wavenumber three-vector and
\begin{equation}
\hat{\boldsymbol{v}}(\boldsymbol{k}) \equiv 
\int_V \boldsymbol{v}(\boldsymbol{x})
e^{-2 \pi i \boldsymbol{k}\cdot\boldsymbol{x}}\, d\boldsymbol{x}\,,
\end{equation}
with $V$ being the three-volume of our computational domain. A number
of recent studies have analyzed the scaling of the velocity power
spectrum in the inertial range, that is, in the range in wavenumbers
between the lengthscale of the problem and the scale at which
dissipation dominates. More specifically, \citet{Inoue2011} has
reported evidences of a Kolmogorov $k^{-5/3}$ scaling in a
freely-decaying MHD turbulence, but has not provided a systematic
convergence study of the spectrum. Evidences for a $k^{-5/3}$ scaling
were also found by \citet{Zhang09}, in the case of the
kinetic-energy spectrum, which coincides with the velocity
power-spectrum in the incompressible case. Finally,
\citet{Zrake2011a} has performed a significantly more systematic
study for driven, transonic, MHD turbulence, but obtained only a very
small (if any) coverage of the inertial range.

\begin{figure}
    \includegraphics[width=\columnwidth]{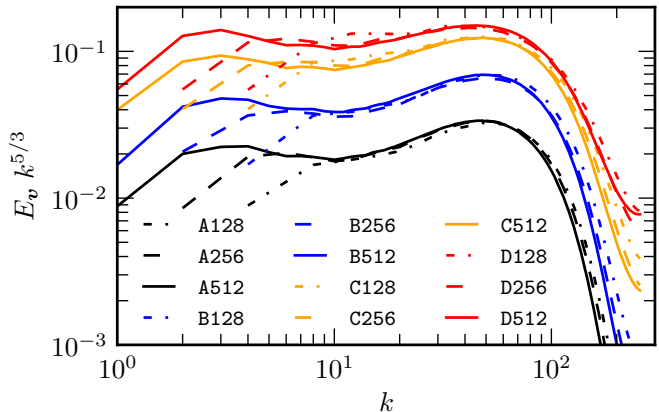}
  \caption{Power spectra of the velocity field.  Different lines refer
    to the three resolutions used and to the different values of the
    driving force. The spectra are scaled assuming a $k^{-5/3}$ law.}
\label{fig:vel_spectrum}
\end{figure}

The time-averaged velocity power spectra computed from our simulations
are shown in Fig.~\ref{fig:vel_spectrum}. Different lines refer to the
three different resolutions used, $128^3$ (dash-dotted), $256^3$ (dashed)
and $512^3$ (solid lines), and to the different values of the driving
force. To highlight the presence and extension of the inertial range, the
spectra are scaled assuming a $k^{-5/3}$ law, with curves at different
resolutions shifted of a factor two or four, and nicely overlapping with
the high-resolution one in the dissipation region. Clearly,
simulations at higher resolutions would be needed to have power-spectra
which are more accurate and with larger inertial ranges, but overall,
Fig.~\ref{fig:vel_spectrum} convincingly demonstrates the good
statistical convergence of our code and gives a strong support to the
idea that the \emph{key} prediction of the Kolmogorov model
(K41)~\citep{Kolmogorov1991a} carries over to the relativistic case.
Indeed, not only does the velocity spectrum for our subsonic model
\texttt{A} shows a region, of about a decade in length,
compatible with a $k^{-5/3}$ scaling, but this continues to be
the case even as we increase the forcing and enter the regime of
relativistic supersonic turbulence with model \texttt{D}. In this
transition, the velocity spectrum in the inertial range, the range of
lengthscales where the flow is scale-invariant, is simply ``shifted
upwards'' in a self-similar way, with a progressive flattening of the
bottleneck region, the bump in the spectrum due to the non-linear
dissipation introduced by our numerical scheme. Steeper or
shallower scalings, such as the Burgers one, $k^{-2}$, or a
$k^{-4/3}$ one, are also clearly incompatible with our data. 

These results have been confirmed in a preliminary study where
we pushed our resolution for model D, the most extreme one, to $1024^3$.

All in all, this is one of our main results: the velocity power spectrum
in the inertial range is \emph{universal}, that is, insensitive to
relativistic effects, at least in the subsonic and mildly supersonic
cases. Note that this does \emph{not} mean that the Kolmogorov theory is
directly applicable to relativistic flows. We point out that the velocity
power spectrum is \emph{not} equal to the kinetic energy density in
Fourier space, as in the classical incompressible case. This is because
of the corrections to the expression of the kinetic energy due to the
fluid compressibility (which is not zero) and the Lorentz factor (we
recall that the relativistic kinetic energy is $T = \rho W(W-1) \simeq
\frac{1}{2}\rho v^2 + \mathcal{O}(v^4)$). For this reason, the
interpretation of the velocity power spectrum requires great care.
Finally we note that already in the Newtonian turbulence the velocity
power-spectrum is known to have large deviations from the $k^{-5/3}$
scalings for highly supersonic flows. In particular \citet{Kritsuk2007}
reported spectra with scaling close to the Burgers one. Similar
deviations could also manifest themselves in the relativistic case for
higher values of the Mach number, but these regimes are currently
not-accessible by our code.
\section{Intermittency}
Not all of the information about relativistic turbulent flows is
contained in the velocity power spectrum. Particularly important in a
relativistic context is the intermittency of the velocity field, that
is, the local appearance of anomalous, short-lived flow features,
which we have studied by looking at the parallel-structure functions
of order $p$
\begin{equation}\label{eq:structure.function}
  S^\parallel_p(r) \equiv \big\langle |\delta_r v|^p \big\rangle,
  \quad
  \delta_r v = \big[
    \boldsymbol{v}(\boldsymbol{x}+\boldsymbol{r}) -
    \boldsymbol{v}(\boldsymbol{x})\big] \cdot \frac{\boldsymbol{r}}{r}
\end{equation}
where $\boldsymbol{r}$ is a vector of length $r$ and the average is
over space and time.

\begin{figure}
  \includegraphics[width=\columnwidth]{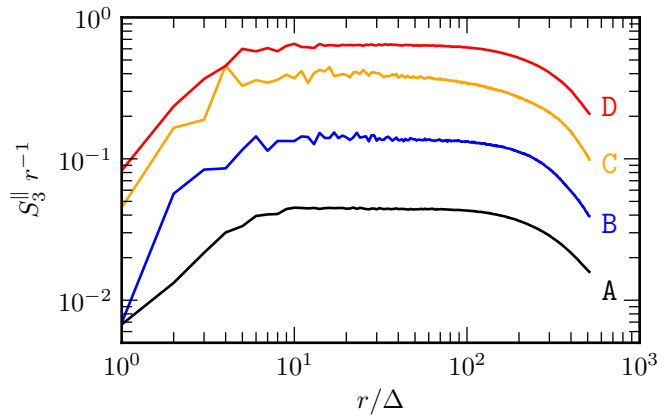}
  \caption{\label{fig:S3} Compensated, third-order, parallel structure
    function computed for the different models as functions of $r /
    \Delta$. Note the very good match with the classical
    $S_3^\parallel \sim r$ behaviour.}
\end{figure}

Figure~\ref{fig:S3} reports the compensated, third-order, parallel
structure function, $S_3^\parallel$, as functions of $r / \Delta$,
where $\Delta$ is the grid spacing. Within the inertial range,
classical incompressible turbulence has a precise prediction: the
Kolmogorov $4/5$-law, for which $\langle (\delta_r v)^3 \rangle =
\frac{4}{5} \epsilon r$, where $\epsilon$ is the kinetic-energy
dissipation rate.  This translates into $S_3^\parallel \sim \epsilon
r$. As shown in the figure, the structure functions are somewhat noisy
at small scales, but are consistent with the classical prediction over
a wide range of lengthscales, with linear fits showing deviations of
$\sim 5\%$, and an increase of $\epsilon$ with the driving force.

Although even in the classical compressible case, the $4/5$-law is not
strictly valid, we can use it to obtain a rough estimate of the turbulent
velocity dissipation rate~\citep{Porter2002}. We find that $\epsilon$, as
measured from $S_3^\parallel$ or directly from $\langle (\delta_r v)^3
\rangle$, grows linearly with the Lorentz factor, in contrast with the
classical theory, where it is known to be independent of the Reynolds
number. This is consistent with the observations that in a relativistic
regime the turbulent velocity shows an exponential decay in
time~\citep{Zrake2011,Inoue2011}, as opposed to the power-law decay seen
in classical compressible and incompressible turbulence. An explanation
for this behaviour might be that, since the inertia of the fluid grows
linearly with the Lorentz factor, an increasingly large rate of energy
injection is needed to balance the kinetic energy losses when the average
Lorentz factor is increased.

\begin{table*}
\caption{Scaling exponents of the parallel structure functions computed
using the ESS technique and analytical predictions from the K41, SL and
Burgers models.}
\label{tab:structure}
\begin{center}
\scriptsize{
\begin{tabular}{lcccccccccc}
\toprule
Model & $\zeta^\parallel_1$ & $\zeta^\parallel_2$ & $\zeta^\parallel_3$ &
$\zeta^\parallel_4$ & $\zeta^\parallel_5$ & $\zeta^\parallel_6$ &
$\zeta^\parallel_7$ & $\zeta^\parallel_8$ & $\zeta^\parallel_9$ &
$\zeta^\parallel_{10}$ \\
\hline
K41 & $0.33$ & $0.67$ & $1$ & $1.33$ & $1.67$ & $2$ & $2.33$ & $2.67$ & $3$ & $3.33$ \\
SL & $0.36$ & $0.70$ & $1$ & $1.28$ & $1.54$ & $1.78$ & $2.00$ & $2.21$ & $2.41$ & $2.59$ \\
Burgers & $0.41$ & $0.74$ & $1$ & $1.21$ & $1.39$ & $1.56$ & $1.70$ & $1.84$ & $1.96$ & $2.08$ \\
\hline
\texttt{A512} & $0.37 \pm 0.01$ & $0.70 \pm 0.02$ &
$1 \pm 0.02$ & $1.27 \pm 0.03$ & $1.51 \pm 0.02$ & $1.72 \pm 0.03$ &
$1.89 \pm 0.04$ & $2.04 \pm 0.04$ & $2.17 \pm 0.03$ & $2.27 \pm 0.02$ \\
\texttt{B512} & $0.36 \pm 0.01$ & $0.70 \pm 0.03$ & $1 \pm 0.04$ & $1.27 \pm 0.05$ & $1.50 \pm 0.07$ & $1.70 \pm 0.08$ & $1.86 \pm 0.12$ & $1.99 \pm 0.16$ & $2.10 \pm 0.21$ & $2.18 \pm 0.26$ \\
\texttt{C512} & $0.37 \pm 0.01$ & $0.70 \pm 0.02$ & $1 \pm 0.03$ & $1.26 \pm 0.04$ & $1.48 \pm 0.05$ & $1.68 \pm 0.07$ & $1.84 \pm 0.09$ & $1.98 \pm 0.11$ & $2.09 \pm 0.13$ & $2.19 \pm 0.16$ \\
\texttt{D512} & $0.38 \pm 0.005$ & $0.71 \pm 0.01$ & $1 \pm 0.03$ & $1.25 \pm 0.03$ & $1.46 \pm 0.05$ & $1.64 \pm 0.07$ & $1.79 \pm 0.09$ & $1.92 \pm 0.11$ & $2.04 \pm 0.14$ & $2.14 \pm 0.16$ \\
\toprule
\end{tabular}
}
\end{center}
\vskip 1.0cm
\end{table*}

The scaling exponents of the parallel structure functions,
$\zeta^\parallel_p$ have been computed up to $p=10$ using the
extended-self-similarity (ESS) technique~\citep{Benzi1993} and are
summarized in Table~\ref{tab:structure}. The errors are estimated by
computing the exponents without the ESS or using only the data at the
final time. We also show the values as computed using the classical K41
theory, as well as using the estimates by She and Leveque
(SL)~\citep{She1994} for incompressible, \ie $\zeta^\parallel_p =
\frac{p}{9} + 2 - 2 (\frac{2}{3})^{p/3}$, and shock-dominated,
\ie $\zeta^\parallel_p = \frac{p}{9} + 1 - (\frac{1}{3})^{p/3}$
\citep{Boldyrev2002}, turbulence.

\begin{figure}
  \includegraphics[width=\columnwidth]{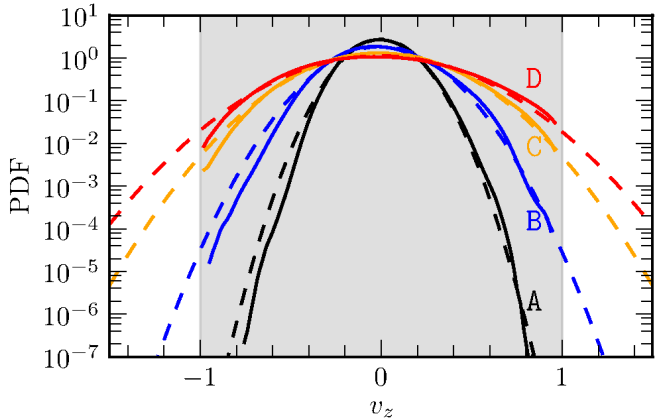}
  \caption{\label{fig:velpdf}PDFs of the velocity $v_z$ for the
    different models considered (solid lines). As the forcing is
    increased, the PDFs flatten, while constrained to be in $(-1,1)$
    (shaded area). Increasingly large deviations from Gaussianity
    (dashed lines) appear in the relativistic regime.}
\end{figure}

Not surprisingly and as also observed in the classical case for high Mach
number flows \citep{Kritsuk2007}\footnote{Note however that
\citet{Kritsuk2007} also find significant deviations in
$\zeta_3^\parallel$ from one, which we do not observe.}, as the flow
becomes supersonic, the high-order exponents tend to flatten out and be
compatible with the Burgers scaling, as the most singular velocity
structures become two-dimensional shock waves.  $\zeta_2^\parallel$,
instead, is compatible with the She-Leveque model even in the supersonic
case. This is consistent with the observed scaling of the velocity power
spectrum, which presents only small intermittency corrections to the
$k^{-5/3}$ scaling. Previous classical studies of weakly
compressible~\citep{Benzi2008} and weakly supersonic
turbulence~\citep{Porter2002} found the scaling exponents to be in very
good agreement with the ones of the incompressible case and to be well
described by the SL model. This is very different from what we observe
even in our subsonic model \texttt{A}, in which the exponents are
significantly flatter than in the SL model, suggesting a stronger
intermittency correction. This deviation is another important result of
our simulations.

One non-classical source of intermittency is the genuinely relativistic
constraint that the velocity field cannot be Gaussian as the PDFs must
have compact support in $(-1, 1)$. This is shown by the behaviour of
the PDFs of $v_z$ and plotted as solid lines in the shaded area of
Fig.~\ref{fig:velpdf}. Clearly, as the Lorentz factor increases, the PDFs
become flatter and, as a consequence, the velocity field shows larger
deviations from Gaussianity (dashed lines). Stated differently,
relativistic turbulence is significantly more intermittent than its
classical counterpart.
\section{Conclusions}
Using a series of high-order direct numerical simulations of driven
relativistic turbulence in a hot plasma, we have explored the
statistical properties of relativistic turbulent flows with average
Mach numbers ranging from $0.4$ to $1.7$ and average Lorentz factors
up to $1.7$. We have found that relativistic effects enhance
significantly the intermittency of the flow and affect the high-order
statistics of the velocity field. Nevertheless, the low-order
statistics appear to be universal, \ie independent from the Lorentz
factor, and in good agreement with the classical Kolmogorov theory.

In the future we plan to pursue a more systematic investigation
of the properties of relativistic turbulent flows at higher resolution.

%
\acknowledgments
We thank M.A. Aloy, P. Cerd\'a-Dur\'an, A. MacFadyen, M. Obergaulinger
and J. Zrake for discussions. The calculations were performed on the
clusters at the AEI and on the SuperMUC cluster at the LRZ. Partial
support comes from the DFG grant SFB/Transregio~7 and by ``CompStar'', a
Research Networking Programme of the ESF.

\end{document}